\documentclass[preprint2]{aastex}

\usepackage{natbib}

\slugcomment{To appear in the Astrophysical Journal}

\shorttitle{BLAST Survey in Vela-D}
\shortauthors{Olmi et al.}

\begin{document}


\title{The BLAST Survey of the Vela Molecular Cloud: \\
Dynamical Properties of the Dense Cores in Vela-D}       

\author{Luca Olmi,\altaffilmark{1,2,\dag}         %
	Daniel Angl\'es-Alc\'azar,\altaffilmark{3}             %
	Massimo De Luca,\altaffilmark{4}              %
	Davide Elia,\altaffilmark{5}                %
	Teresa Giannini,\altaffilmark{6}            
	Dario Lorenzetti,\altaffilmark{6}          
	Fabrizio Massi,\altaffilmark{2}                
	Peter G. Martin,\altaffilmark{7,8} 
	Francesco Strafella\altaffilmark{9} }          

\altaffiltext{1}{University of Puerto Rico, Rio Piedras Campus, Physics Department, Box 23343, 
UPR station, San Juan, Puerto Rico}

\altaffiltext{2}{Osservatorio Astrofisico di Arcetri - INAF, Largo
E. Fermi 5, I-50125, Firenze, Italy.}

\altaffiltext{3}{University of Arizona, Department of Physics, 1118 E. Fourth Street,
PO Box 210081, Tucson, AZ 85721}

\altaffiltext{4}{LERMA-LRA, UMR 8112 du CNRS, Observatoire de Paris, \'Ecole Normale
Sup\'erieure, UPMC \& UCP, 24 rue Lhomond, 75231 Paris Cedex 05, France}

\altaffiltext{5}{Istituto di Fisica dello Spazio Interplanetario - INAF, 
via Fosso del Cavaliere 100, I-00133 Roma, Italy.}

\altaffiltext{6}{Osservatorio Astronomico di Roma - INAF, Via Frascati 33, I-00040
Monteporzio Catone, Roma, Italy.}

\altaffiltext{7}{Canadian Institute for Theoretical Astrophysics, University of Toronto, 60
St. George Street, Toronto, ON M5S~3H8, Canada}

\altaffiltext{8}{Department of Astronomy \& Astrophysics, University of Toronto, 50 St.
George Street, Toronto, ON  M5S~3H4, Canada}

\altaffiltext{9}{Dipartimento di Fisica, Universit\'a del Salento, CP 193,
I-73100 Lecce, Italy.}

\altaffiltext{\dag}{\url{olmi.luca@gmail.com, olmi@arcetri.astro.it}}

\clearpage

\begin{abstract}
The Vela-D region, according to the nomenclature given by \citet{mur:may},
 of the star forming complex known as the Vela Molecular Ridge (VMR), has been
recently analyzed in details by \citet{olmi2009}, who studied the physical properties of 
141 pre- and proto-stellar cold dust cores, detected by the ``Balloon-borne Large-Aperture
Submillimeter Telescope'' (BLAST) during a much larger (55~deg$^2$) Galactic Plane survey 
encompassing the whole VMR. This survey's primary goal was to identify
the coldest, dense dust cores possibly associated with the earliest phases of star formation.
In this work, the dynamical state of the Vela-D cores is analyzed.
Comparison to dynamical masses of a sub-sample of the Vela-D cores 
estimated from the $^{13}$CO survey of \citet{elia2007}, is 
complicated by the fact that the $^{13}$CO linewidths
are likely to trace the lower density intercore material, in addition
to the dense gas associated with the compact cores observed by BLAST.
In fact, the total internal pressure of these cores, if estimated using the 
$^{13}$CO linewidths, appears to be higher than the cloud ambient pressure. If this were the case,
then self-gravity and surface pressure would be insufficient to bind these cores
and an additional source of external confinement (e.g., magnetic field pressure) would be required. 
However, if one attempts to scale down the $^{13}$CO linewidths, according to the observations
of high-density tracers in a small sample of sources, then most proto-stellar cores would result
effectively gravitationally bound.
\end{abstract}


\section{INTRODUCTION }

Pre-stellar cores represent a very early stage of the star formation (SF)
process, before collapse results in the formation of a central protostar.
The physical and dynamical properties of these cores can reveal important clues about their
nature.
More than a thousand new pre- and proto-stellar cores have been identified by BLAST,
the Balloon-borne Large-Aperture Submillimeter Telescope \citep{pascale2008},
during its second long duration balloon (LDB) science flight in 2006. BLAST detected
these cold cores in a $\sim 50\,$deg$^2$ map of the Vela Molecular Ridge (VMR) 
\citep{netterfield2009}, and was capable of constraining the temperatures of the detected sources 
using its three-band photometry (250, 350, and 500\,\micron) near the peak of the
cold core SED.

The VMR (\citealp{mur:may}, \citealp{Lis92}) is a giant molecular cloud 
complex within the galactic plane, in the area
$260 \degr \la l \la 272 \degr$ and $-2 \degr \la b \la 3 \degr$,
hence located outside the solar circle. Its main properties have been recently revisited
by \citet{netterfield2009} and \citet{olmi2009}. 
In particular, \citet{olmi2009} used both BLAST and archival data to determine the  
spectral energy densities (SEDs) and the physical parameters
of each source detected by BLAST in the smaller region of Vela-D, where observations
from IR (\citealp{gianni07}) to millimeter wavelengths (\citealp{massi07}) were already available. 

In this work we will use the spectral line data
of \citet{elia2007} to perform an analysis of the dynamical state of the cores.
In Section~\ref{sec:dyn} we discuss which cores are gravitationally bound. Then
in Section~\ref{sec:pressure} we discuss the effects of pressure.
We draw our conclusions in Section~\ref{sec:concl}.

\section{DYNAMICAL STATE OF THE CORES }
\label{sec:dyn}

\subsection{Core Size Distribution}
\label{sec:massize}
\begin{figure}
\hspace*{-1cm}
\includegraphics[width=10.5cm]{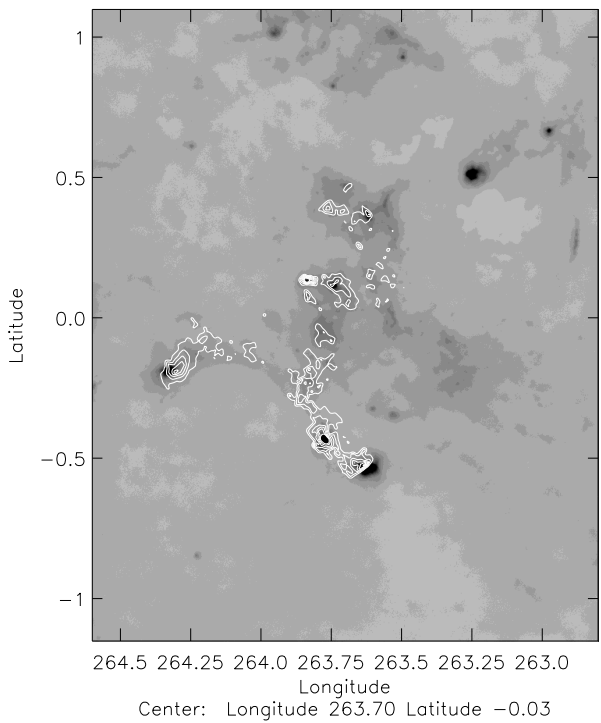}
\caption{
Gray-scale image shows the BLAST 250\,\micron\  map of Vela-D, with
galactic coordinates in degrees. Solid contours represent the $^{13}$CO$(2-1)$
emission (from \citealp{elia2007}), integrated between 2\,km~s$^{-1}$ and 15\,km~s$^{-1}$.
The first contour is at 2\,K\,km~s$^{-1}$, and contour spacing is $\simeq 7\,$K\,km~s$^{-1}$.
  }
\label{fig:overlay}
\end{figure}

The cores in Vela-D have  a similar range of 
sizes ($\langle R_c \rangle = 0.09 \pm 0.02$~pc) to those in the overall VMR 
($\langle R_c \rangle = 0.07 \pm 0.02$~pc), for comparable range of masses.  
Please note that although we use the term core size, or core radius, what we actually
measure with BLAST is an ``effective core width'', set by the power-law distribution
of the column density in the core and by the BLAST beam.
The small range of
sizes for a large range of masses in both VMR and Vela-D is similar to what Enoch et al. (2008)
found for the proto-stellar cores in Serpens, Ophiucus and, to a lesser
degree, in Perseus. 

We also evaluated  the ratios of angular deconvolved size (from BLAST) to beam size,
$\theta_{\rm dec}/\theta_{\rm mb}$.  Power-law density profiles in pre-stellar cores have 
received both a theoretical (see for example the discussions in \citealp{young2003} and 
\citealp{mckee2007}) and observational support (e.g., \citealp{ward1999}, \citealp{andre2000} and
references therein). Assuming that also the Vela-D cores can be 
described by power-law density profiles, then according to \citet{young2003}
$\theta_{\rm dec}/\theta_{\rm mb}$ should depend only on the index of the power law, and not on the
distance of the source. We find $\theta_{\rm dec}/\theta_{\rm mb}$ to have a
median value of  1.41 for Vela-D, corresponding 
to a density power-law exponent $p\simeq 1.6$  \citep{young2003}.     
This value for the power-law exponent, similar to that found by \citet{enoch2008} in Ophiucus,
is consistent with mean $p$-values 
found from radiative transfer modeling of Class 0 and Class I envelopes 
\citep{young2003}.

\subsection{Bonnor-Ebert Analysis }
\label{sec:BEmass}

In this section we investigate the dynamical state of starless and
proto-stellar cores, in order to determine whether or not the
BLAST starless cores in Vela-D are pre-stellar,
i.e. they will collapse and form one or more stars. 
We first analyze the dynamical state of the cores by calculating
the critical Bonnor-Ebert mass, $M_{\rm BE}$. In this analysis we assume that
all cores are spherical in shape, as our angular resolution
at the distance of Vela-D is not sufficient to determine with enough
accuracy the ellipticity of the cores.
Because elongated cores can be considerably more stable than expected from a
pure spherical Bonnor-Ebert model (e.g., \citealp{sp2004}), then we may be overestimating the
non-stability of some cores.
Furthermore, it should be noted that even if the core mass is larger than
the Bonnor-Ebert mass, this condition alone does not guarantee that
the core is truly gravitationally unstable, as we are neglecting any other
non-thermal support.  Other sources of support are discussed in the next sections.

%
%
\begin{figure}
\hspace*{-1cm}
\includegraphics[width=8.5cm]{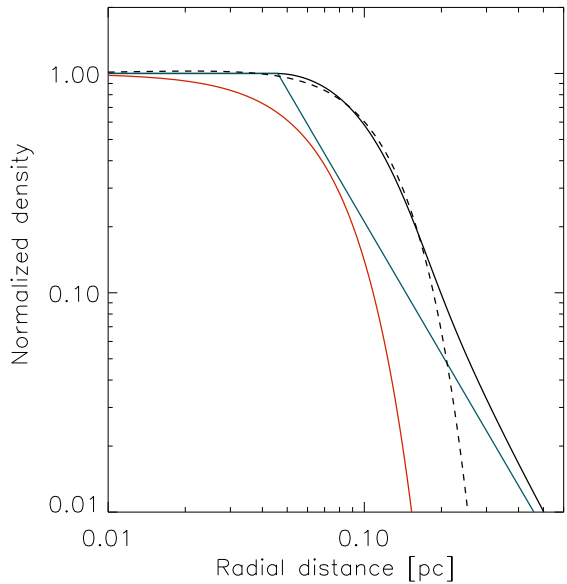}
\caption{
Example showing how $R_{\rm BE}$ is estimated. The red solid line represents the profile of
the BLAST 250\,\micron\ beam at the distance of Vela-D. The blue solid line shows the approximated
Bonnor-Ebert density profile, from which the simulated $R_{\rm BE}$ is calculated.
The black solid line represents the convolution of the beam and density profiles, and the dashed
black line shows the Gaussian best-fit to the convolved profile, from which the source
simulated radius, $R^{\rm sim}$ (HWHM), is determined.
  }
\label{fig:berad}
\end{figure}

The Bonnor-Ebert mass, $M_{\rm BE}$, can be determined from the 
observables, i.e. the core radius and temperature:
\begin{equation}
M_{\rm BE} [M_\odot] = 1.8 R_{\rm BE} [{\rm pc}] \, T [{\rm K}]
\label{eq:bemass}
\end{equation}
%
where $T$ is the core temperature, as determined from the SED fits \citep{olmi2009},
and $R_{\rm BE}$ represents the core radius at which the internal and ambient pressures are equal,
for a Bonnor-Ebert sphere in hydrostatic equilibrium.

$R_{\rm BE}$ can be estimated as $R_{\rm BE} = f \, R_{\rm dec}$,  
where $R_{\rm dec}$ is the measured linear deconvolved half-width at half-maximum 
and $f$ is a numerical factor that has been determined as follows. 
For each core the approximated Bonnor-Ebert density profile for an isothermal sphere,
flat at the core center\footnote{The core center is defined here as the region where the normalized radius
$r_{\rm n} < 1$, with $r_{\rm n} = r \, \sqrt{4\pi G m_{\rm av} \rho_\circ / (k T)}$, where $m_{\rm av}$
is the mass of the average gas particle and $\rho_\circ$ is the density at the center of the core.} 
and falling off with $r^{-2}$ to the outside (see Figure~\ref{fig:berad}),
is generated. The density profile is  obtained using the temperature and average 
density estimated for each core, 
and the theoretical radius, $R_{\rm BE}$, of this sphere is estimated. This profile
is then convolved with the BLAST beam at 250\,\micron\ and then fitted with a Gaussian, 
obtaining a simulated ``observed'' radius $R^{\rm sim}$ (defined as the half-width at half-maximum
of the fitting Gaussian). Finally,
this simulated radius is deconvolved, obtaining $R_{\rm dec}^{\rm sim}$,
and then compared with $R_{\rm BE}$. Their ratio is then the factor 
$f = R_{\rm BE} / R_{\rm dec}^{\rm sim}$ specific to that source. The factor $f$ 
can then be used to determine $R_{\rm BE}$ from the {\it measured} radius, $R_{\rm dec}$, thus
keeping into account any mismatch between $R_{\rm dec}^{\rm sim}$ and $R_{\rm dec}$.

At the distance of Vela-D, and for the range of core temperatures found in this cloud,
we find an average value $\langle f \rangle = 2.55\pm0.14$. Thus, from the observed $R_{\rm dec}$
and core temperature, the value of $M_{\rm BE}$ can be estimated.
To further rid the sample of unrealistically
small core sizes we also eliminated all data for which the beam radius
exceeded the deconvolved source radius. See \citet{jijina1999}  and \citet{enoch2007} 
for further discussion on the effects of instrumental resolution.

%
%
\begin{figure}
\hspace*{-1cm}
\includegraphics[width=9.0cm]{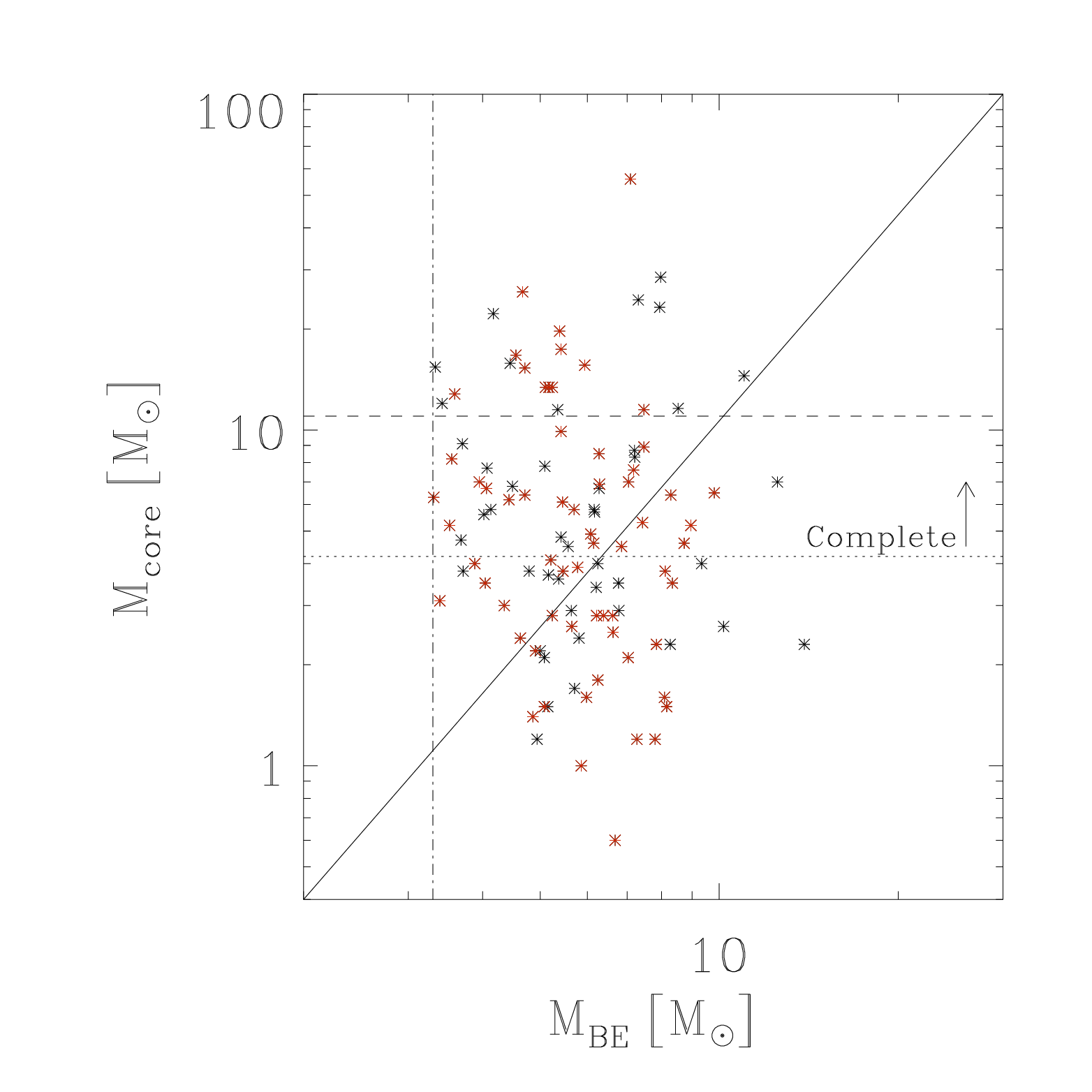}
\caption{
Total core mass, derived from the SED fits, vs. the Bonnor-Ebert mass.
Red symbols are for starless cores, black symbols for proto-stellar cores.
The median value of $M_{\rm BE}$ is  $5.7 \, M_\odot$.
The solid line corresponds to $M_{\rm core} = M_{\rm BE}$.
The two horizontal lines correspond to the completeness limits for sources with $T>12\,$K 
(dotted line) and $T>10\,$K (dashed line), respectively \citep{olmi2009}.
The vertical dot-dashed line correspond to the estimated minimum measurable Bonnor-Ebert mass,
$M_{\rm BE}^{\rm min}$.
  }
\label{fig:bemass}
\end{figure}

%
%

Figure~\ref{fig:bemass} plots the
total core mass, $M_{\rm core}$, also determined from the SED fits, vs. the Bonnor-Ebert mass 
for all BLAST cores in Vela-D that have well-determined SED fits. We also plot the
completeness limits and a vertical -dashed line corresponding to the estimated minimum 
measurable Bonnor-Ebert mass, calculated as;
\begin{equation}
M_{\rm BE}^{\rm min} [M_\odot] = 1.8 R_{\rm BE}^{\rm min} [{\rm pc}] \, T_{\rm compl} [{\rm K}]
\end{equation}
where $T_{\rm compl}=12\,$K, and $R_{\rm BE}^{\rm min} = \langle f \rangle \, R_{\rm beam}$, where
$R_{\rm beam}$ corresponds to the beam linear half-width at half-maximum at the distance of Vela-D.

The cores with $M_{\rm core} \ge M_{\rm BE}$ represent sources which, {\it in the absence of
further internal support} (turbulent, magnetic, etc.), are gravitationally unstable and will
collapse (or are already collapsing) in the presence of external perturbations. Without
further knowledge of the internal structure of these cores we {\it cannot} yet establish
whether they are collapsing  or will collapse to form stars.
On the other hand, the cores with $M_{\rm core} < M_{\rm BE}$ represent sources
which, assuming only thermal support as in Bonnor-Ebert spheres, are {\it stable} against 
external perturbations.  Any further internal support will thus add
to their stability against gravitational collapse, and might eventually lead to the core
disruption.


We find a median $M_{\rm core}/M_{\rm BE} = 0.9$ for all sources in Vela-D,
i.e. very near the critical value, and indicating that we have about the same number 
of sources with $M_{\rm core} < M_{\rm BE}$ and $M_{\rm core} > M_{\rm BE}$. 
However, if we restrict ourselves to mass ranges for which we are progressively more 
complete ($M_{\rm core} \ga 4\, M_\odot$, see Table~3 of \citealp{olmi2009}),
our sources exceed their corresponding 
Bonnor-Ebert mass by a median factor of $M_{\rm core}/M_{\rm BE} = 1.5 - 3.3$.
Figure~\ref{fig:bemass} shows very little difference between starless and proto-stellar cores,
and we can also see that the core mass range where the critical 
value $M_{\rm core} = M_{\rm BE}$ is crossed is $M_{\rm core} \sim 3 - 8 \, M_\odot$.
In this range we are only partially complete (see \citealp{netterfield2009} 
and \citealp{olmi2009}) and thus more sensitive measurements are needed in order to 
confirm that this is an evolutionary effect.

\subsection{Virial Analysis}
\label{sec:virial}

\subsubsection{Determination of Virial Masses}
\label{sec:virialmass}

We now investigate the dynamical state of the Vela-D cores by estimating
their virial masses. The comparison of dust-derived masses
to molecular line observations gives a more robust method of determining
if the cores are gravitationally bound. In fact, while our determination of the
dust masses provide an estimate of the gravitational potential energy,
the molecular linewidths constitute a measure of the internal energy of the
cores, provided that the selected molecular lines are truly tracers of the 
cores observed at (sub)millimeter wavelengths. This is typically the case
with molecular probes of the interior of pre- and proto-stellar cores,
such as NH$_3$ and N$_2$H$^+$, whereas more abundant and diffuse 
molecules, such as $^{13}$CO described below, will provide only an upper limit to the
internal energy of the cores.

Here we will be using the $^{13}$CO$(2-1)$ data of \citet{elia2007},
but instead of using the catalog of these authors (obtained by
CLUMPFIND, \citealp{williams1994}) to associate a molecular clump to a BLAST core, we use the
coordinates of the BLAST objects (when they fall within the $^{13}$CO map of \citealp{elia2007}) 
to find the corresponding $^{13}$CO$(2-1)$
{\it spectrum} along the line of sight. This procedure has two advantages:
first, we avoid associating the same catalog clump (generally somewhat
more massive and larger in size) to multiple BLAST cores, and in addition we can directly
perform a Gaussian fit of the $^{13}$CO$(2-1)$ line without having to rely
on the ability of CLUMPFIND to separate adjacent (in space and velocity) clumps.

%
We selected only BLAST cores with corresponding well-determined 
$^{13}$CO$(2-1)$ linewidths (a total of 35 sources). 
In the spectra of four sources  
we find multiple $^{13}$CO velocity components along the line-of-sight to the BLAST core. 
However, in three of these sources we observe only two velocity components, and their linewidths are 
very similar or they differ by less than 50\,\%. Thus, as long as the linewidths are not too different,
it does not matter which of the two $^{13}$CO$(2-1)$ velocity components is actually associated with 
the BLAST core.
In the remaining source 
we find multiple velocity components with linewidths ranging from 0.3\,km~s$^{-1}$ 
to 1.8\,km~s$^{-1}$, and because we cannot 
know with certainty which one is actually associated with
the BLAST core, we did not use it for the determination of the virial mass.

%
%
\begin{figure}
\centering
\includegraphics[width=8.0cm]{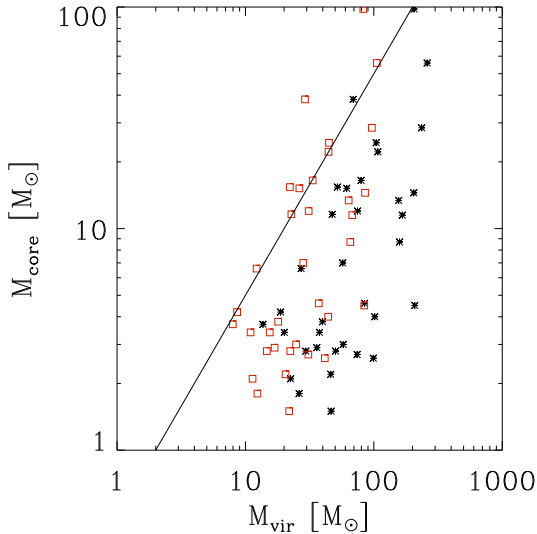}
\caption{
Total core mass, derived from the SED fits, vs. the virial mass,
calculated using the velocity linewidth derived from
the $^{13}$CO$(2-1)$ data of \citet{elia2007}.
The solid line indicates the minimum $M_{\rm core} = 0.5 M_{\rm vir} = 0.5 M_{\rm BET}$ for which the cores
should be self-gravitating.
Black asterisks represent the results when no scale factor is applied to the $^{13}$CO$(2-1)$
linewidths, whereas the red open squares show the resulting
parameters when the linewidths are scaled (with the 1.6 factor, see text) 
and thus are thought to represent more closely
the core internal velocity dispersion. For the sake of simplicity,
both starless and proto-stellar cores are represented by the same symbols.
  }
\label{fig:virial}
\end{figure}

We estimate the virial mass of the cores as \citep{maclaren1988}: 
%
\begin{equation}
M_{\rm vir}{\rm [M_\odot]} = 210 \, R_{\rm dec}{\rm [pc]} \, (\Delta V {\rm [km~s^{-1}]})^2
\end{equation}
where $\Delta V$ represents the total linewidth of the mean particle, and is
calculated as the sum of the thermal and turbulent components:
%
\begin{equation}
\Delta V^2 = \Delta V_{\rm 13CO}^2 + kT8 \ln 2 \, \left ( \frac{1}{m_{av}} -
\frac{1}{m_{\rm 13CO}} \right )
\end{equation}
where $\Delta V_{\rm 13CO}$ is the $^{13}$CO$(2-1)$ line FWHM, $m_{av}=2.3$~amu is
the mean molecular weight and $m_{\rm 13CO}=29$~amu is the molecular weight of
$^{13}$CO.

Figure~\ref{fig:virial} plots the
total core mass $M_{\rm core}$  vs. the virial mass $M_{\rm vir}$ 
for all the BLAST cores in Vela-D selected as described above. 
The solid line in this figure represents the self-gravitating limit
defined by $K = -U$, where $K$ is the kinetic and $U$ the gravitational 
potential energy, corresponding to $M_{\rm core} = 0.5 \, M_{\rm vir}$.
We note that one could expand the definition of Bonnor-Ebert mass to include the 
effects of turbulence, thus writing Eq.~(\ref{eq:bemass}) as:
\begin{equation}
M_{\rm BET} [M_\odot] = 1.8 R_{\rm BE} [{\rm pc}] \, T_{\rm eff} [{\rm K}]
\label{eq:bemasst}
\end{equation}
where the effective temperature, $T_{\rm eff}$, is defined as the temperature that 
describes the total line width, i.e.:
\begin{equation}
\Delta V^2 = \frac{k T_{\rm eff} \, 8 \ln 2}{m_{av}} 
\end{equation}
which also gives $T_{\rm eff} \, [{\rm K}] = 50 \, \Delta V^2 \, [{\rm km~s^{-1}}]$.
Eq.~(\ref{eq:bemasst}) then becomes:
\begin{equation}
M_{\rm BET} [M_\odot] = 90 f \, R_{\rm dec} [{\rm pc}] \, \Delta V^2 \, [{\rm km~s^{-1}}] \, ,
\end{equation}
where we have used $R_{\rm BE} = f \, R_{\rm dec}$ (see Section~\ref{sec:BEmass}).
We therefore see that $M_{\rm BET} \simeq M_{\rm vir}$, and the self-gravitating limit corresponds to 
$M_{\rm core} = 0.5 \, M_{\rm vir} \simeq 0.5 \, M_{\rm BET}$.

In Figure~\ref{fig:virial} we can see that nearly all of the 
cores  
lie below the self-gravitating line,
indicating that they are unlikely to be gravitationally bound, in agreement
with the findings of \citet{elia2007} and \citet{massi07}.
However, the $^{13}$CO$(2-1)$ emission may be partially optically thick along
the line of sight of the $^{13}$CO clump centroid
(see \citealp{elia2007}), and even when it is indeed optically thin 
it traces all molecular gas along the 
line of sight. Although, as we discussed earlier, we measure the 
$^{13}$CO linewidths along the line of sight to the BLAST cores and not
toward the clumps centroid of \citet{elia2007}, we
are likely to trace the lower density intercore material in addition
to the dense gas associated with the compact cores observed by BLAST.
We attempt to correct for this effect in the next section.


\subsubsection{Application of a Scaling Factor for the Linewidths}
\label{sec:linewidth}

Recent follow-up observations toward a sample of the Vela-D cores 
(Olmi et al., {\it in preparation}) have allowed us to determine 
the average of the $^{13}$CO$(2-1)$ to N$_2$H$^+(1-0)$ linewidth
ratio, equal to $\simeq 1.6$, in a small sub-sample of 7 cores. 
For comparison, in the Pipe cores Rathborne et al. (2008) find that 
the C$^{18}$O linewidths, in the range $\sim 0.14 - 0.61$~km~s$^{-1}$
\citep{muench2007}, are typically 1.4 times broader than NH$_3$
which, like N$_2$H$^+$, traces only the gas associated with the dense compact core. 
Furthermore, \citet{onishi1999} found that the $^{13}$CO emission
lines from the Pipe cloud were characterized by linewidths of 
$\sim 1$~km~s$^{-1}$. This leads in this case to a typical $^{13}$CO to NH$_3$ 
linewidth ratio of $\simeq 2.8$. 

The difference between the two scale factors may be caused, besides to the 
statistical significance of the two samples of sources, also by 
real dynamical differences between the Vela-D and Pipe regions. 
In fact, both \citet{elia2007} and \citet{massi07} find evidence for 
turbulent activity in Vela-D. In addition, the presence of multiple IRAS embedded clusters, stellar
jets and outflows also suggest that overall Vela-D is in a more advanced evolutionary state 
compared to the Pipe nebula. 

We also note that if we use the NH$_3$ survey of \citet{jijina1999} and
the  $^{13}$CO survey of \citet{vilasboas2000}, the average $^{13}$CO to NH$_3$ linewidth ratios
toward, e.g., the Ophiucus and Taurus clouds would be even larger than that estimated for the Pipe cores.
Therefore, the reader should consider the results obtained by applying the linewidth scale factors
with caution. In the following, we will use the 1.6 scale factor obtained toward Vela-D as our baseline value, but
we will occasionally discuss how the results would change if a larger (typical of other clouds) scale factor
were used.

\subsubsection{Comparison with Previous Results}
\label{sec:prevres}

Comparing Figures~\ref{fig:bemass} and \ref{fig:virial} we note that
the core mass, $M_{\rm core} \simeq 3 - 8 \, M_\odot$, 
at which the critical threshold $M_{\rm core} =  M_{\rm BE}$ is crossed,
is quite lower than the mass at which the
cores appear to become gravitationally bound, which from Figure~\ref{fig:virial}
can be roughly estimated to be $\ga30 \, M_\odot$, when the scale factor
for the linewidths is applied. 
%
Looking at Figure~12 of \citet{olmi2009}, it would be tempting to infer that 
the characteristic mass at which the core mass function  appears to break or depart 
from a power-law is similar with the critical Bonnor-Ebert mass of Figure~\ref{fig:bemass}. 
However, we cannot compare these mass ranges
because of the completeness limit. Thus,
we cannot confirm for Vela-D the results obtained by \citet{lada2008} for the Pipe cores,
where the authors find a consistency between all of these critical mass ranges.

As far as the results of \citet{elia2007} are concerned, we have already seen that we associate
a BLAST core to a single $^{13}$CO$(2-1)$ velocity component, whereas in the clump catalog of 
\citet{elia2007} multiple gas clumps may be associated with a group of dust cores, as detected by \citet{massi07}.
Because of this and also because \citet{elia2007} do not attempt to correct the $^{13}$CO$(2-1)$ linewidths, 
it is not surprising that their virial masses turn out to be a factor $\sim 3 - 40$ larger than the estimated gas mass
of each clump. 

On the other hand, assuming that either N$_2$H$^+$ or NH$_3$ trace the bulk of the
core emission, one can apply the 1.6 scale factor estimated in Section~\ref{sec:linewidth} 
to the $^{13}$CO$(2-1)$ linewidths. In this case we find that $\simeq 35\,$\% of the Vela-D cores
detected by BLAST within the area covered by the $^{13}$CO$(2-1)$ map of \citet{elia2007}
are near or over the threshold to become gravitationally bound, as shown in Figure~\ref{fig:virial}.
If we had used a larger linewidth scale factor, such as the one derived from the Pipe cores,
nearly half of the Vela-D cores would result gravitationally bound,
according to this analysis. However, the exact fraction of bound
cores will be known only when the linewidth of each core will be
individually measured with a high-density tracer.



%
%
\begin{figure}
\hspace*{-1cm}
\includegraphics[width=8.5cm]{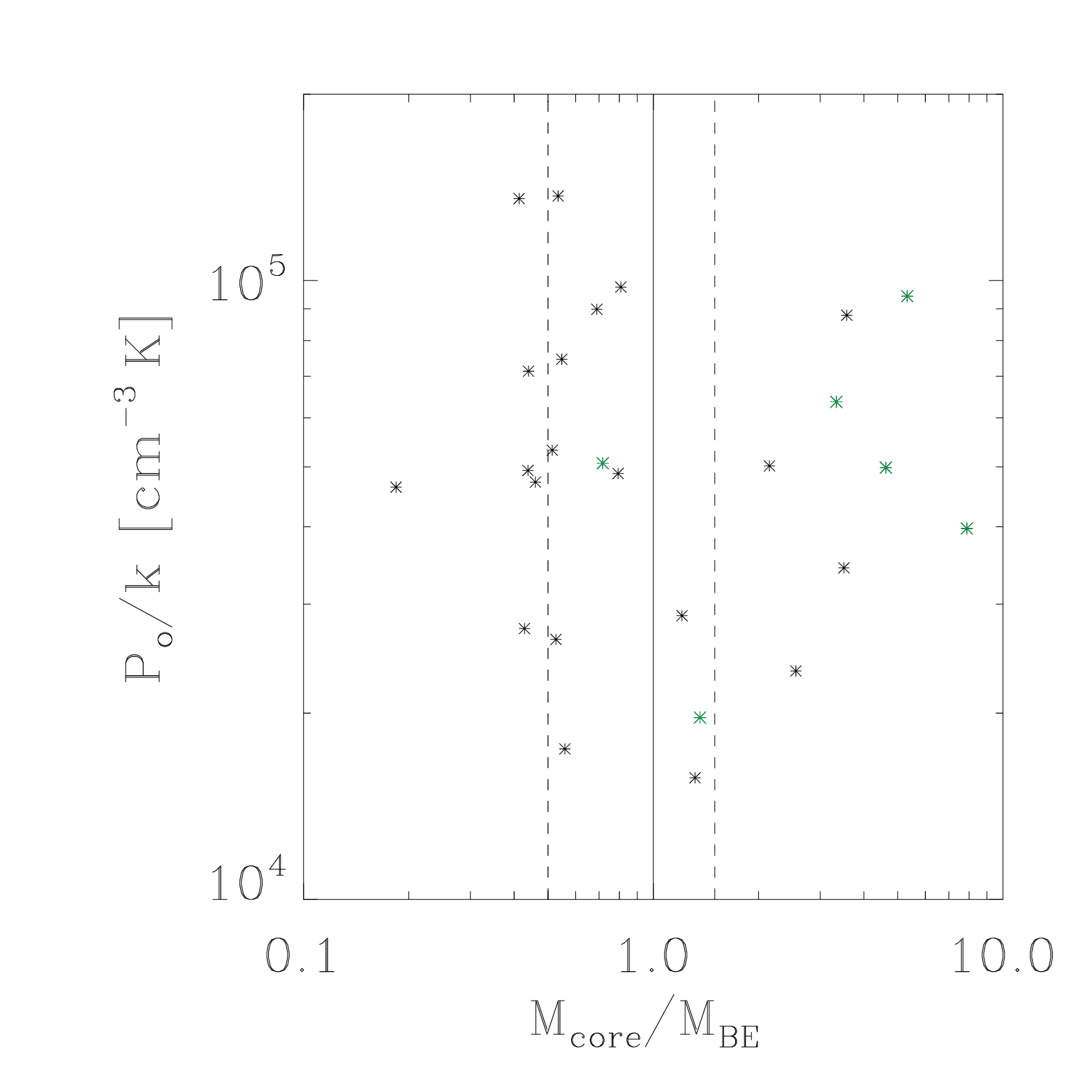}
\hspace*{-1cm}
\includegraphics[width=8.5cm]{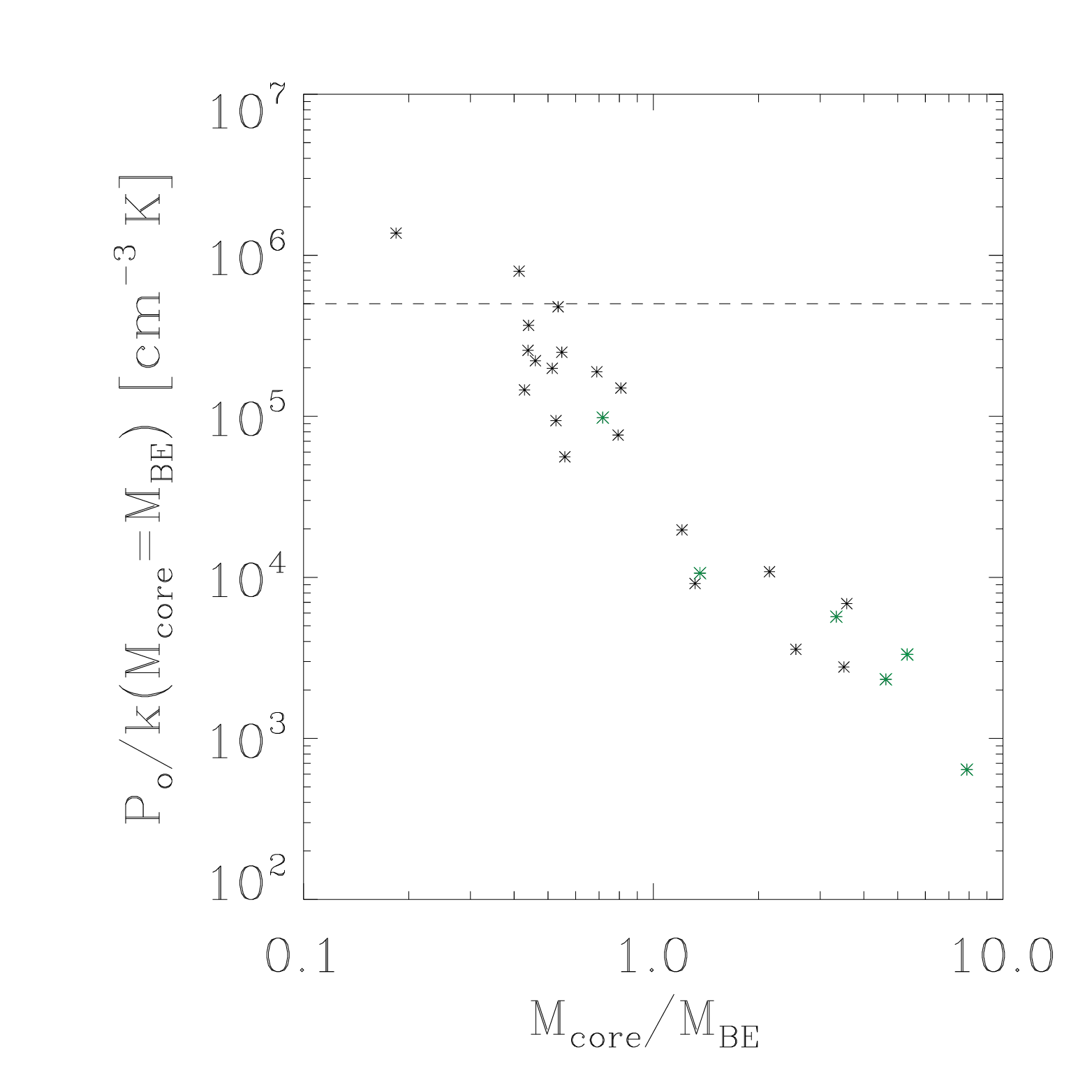}
\caption{
{\it Top.}
Pressure, $P_\circ/k$, at the edge of the Bonnor-Ebert sphere (equal to the ambient pressure) vs.
$M_{\rm core}/M_{\rm BE}$, for all points where a virial mass could be determined.  
The green symbols represent cores with $M_{\rm core} \ge 0.5 \,M_{\rm vir}$
(1.6 scale factor applied, see text; black symbols are for cores with 
$M_{\rm core} < 0.5 \,M_{\rm vir}$).
{\it Bottom.} 
Edge pressure calculated assuming $M_{\rm BE} = M_{\rm core}$ (see text). 
Green and black symbols are as before.
The dashed horizontal line corresponds to the ambient pressure in Vela-D, 
$P_{ext}/k \sim 5 \times 10^5$~cm$^{-3}$~K (see Section~\ref{sec:therm}).
  }
\label{fig:bepress}
\end{figure}

\section{EFFECTS OF PRESSURE}
\label{sec:pressure}

\subsection{Cores Edge Pressure}
\label{sec:BEpress}

To further investigate the dynamical state of the Vela-D cores we now analyze the
effects of thermal and non-thermal pressures on the cores.
First, we compare the core mass with the estimated radius of the Bonnor-Ebert
sphere, for a given edge pressure.  
The Bonnor-Ebert critical mass can be written in terms of its radius and of the pressure,
$P_\circ$, at the edge of the sphere (equal to the ambient pressure) as:
\begin{equation}
M_{\rm BE} [M_\odot] = 0.15 \, (D_{\rm BE} [{\rm pc}])^2 \, (P_\circ/k [{\rm cm^{-3} \, K}])^{1/2}.
\label{eq:bemassrad}
\end{equation}
%
%
In the top panel of Figure~\ref{fig:bepress} we have thus plotted for each core where a measurement of
$M_{\rm vir}$ was available, the edge pressure $P_\circ/k$,
corresponding to a Bonnor-Ebert sphere (with given radius and temperature) in hydrostatic equilibrium,
as a function of the $M_{\rm core}/M_{\rm BE}$ ratio. In addition, we have also plotted the sources
where $M_{\rm core}/M_{\rm vir} \ge 0.5$. 

Given the small number of data points, the interpretation of the top panel of 
Figure~\ref{fig:bepress} is difficult. The number of data points becomes even smaller if we
take into account the fact that the definition of $P_\circ$ is only valid for sources {\it at or near
the Bonnor-Ebert equilibrium}. For this reason we have drawn in Figure~\ref{fig:bepress} two dashed
vertical lines corresponding to $M_{\rm core}/M_{\rm BE} = 0.5, 2$ which enclose all sources
having a mass within a factor of 2 from the corresponding Bonnor-Ebert
mass required to be in hydrostatic equilibrium.
For these points there is a wide distribution of $P_\circ/k$ values, from
$P_\circ/k \sim 10^4$~cm$^{-3}$~K to $\sim 10^5$~cm$^{-3}$~K. However, we tentatively note that most sources
with $M_{\rm core}/M_{\rm BE} < 1$ have a {\it higher} ($\ga 4 \times 10^4$~cm$^{-3}$~K) 
value of the edge pressure, compared to those objects with $M_{\rm core}/M_{\rm BE} > 1$ (three sources 
only, with pressure $\la 3 \times 10^4$~cm$^{-3}$~K). 
This is indeed what one might expect in cores with low density-contrast which are mainly
confined by the external pressure, and not by self-gravity. Clearly, a larger source sample is
needed to confirm that this trend is real.
We also note that most sources with $M_{\rm core}/M_{\rm vir} \ge 0.5$ (with a 1.6 scale factor) also
have $M_{\rm core}/M_{\rm BE} > 1$, a trend which is confirmed when a larger linewidth scale factor is used.

In the bottom panel of Figure~\ref{fig:bepress} we have instead plotted the edge pressure, 
$P_\circ/k$, assuming $M_{\rm BE} = M_{\rm core}$, as a function of the $M_{\rm core}/M_{\rm BE}$ ratio,
estimated as described in Section~\ref{sec:BEmass}. From Equations~(\ref{eq:bemass}) and 
(\ref{eq:bemassrad}) one can see that $P_\circ/k \propto T^4 \, M_{\rm BE}^{-2}$. Therefore,
for cores with $M_{\rm core}/M_{\rm BE} < 1$, the edge pressure calculated as
$P_\circ/k \propto T^4 \, M_{\rm core}^{-2}$ represents the {\it external} pressure required 
to make the core gravitationally {\it unstable}. On the other hand, in the cores where
$M_{\rm core}/M_{\rm BE} > 1$ the quantity $P_\circ/k \propto T^4 \, M_{\rm core}^{-2}$ 
represents the {\it internal} pressure required to make the core {\it stable} against
gravitational collapse.

Since the ambient pressure in the densest parts of the Vela-D region has been estimated to be 
$P_{ext}/k \sim 5 \times 10^5$~cm$^{-3}$~K (see Section~\ref{sec:therm}) the bottom panel
of Figure~\ref{fig:bepress} is suggesting that the cores with $M_{\rm core}/M_{\rm BE} < 1$
are actually likely to undergo gravitational collapse at some point in the future, unless
an additional, non-thermal source of internal support exist. This internal support could in fact be
of turbulent origin, since most of the points with 
$M_{\rm core}/M_{\rm BE} < 1$ also have $M_{\rm core}/M_{\rm vir} < 0.5$ (see Section~\ref{sec:virial}).

On the other hand, the cores with $M_{\rm core}/M_{\rm BE} > 1$ show that the required
internal, thermal pressure to remain in equilibrium is much less than the ambient pressure
in Vela-D. Thus, these cores  are also likely to undergo gravitational collapse, except possibly
the objects with $M_{\rm core}/M_{\rm vir} < 0.5$, which have an additional source of
internal turbulent support.


\subsection{Thermal and non-Thermal Pressure}
\label{sec:therm}

We now specifically consider the ratio of thermal to non-thermal pressure,
$R_p = (a / \sigma_{\rm NT})^2$, for each core; here $a = [k T / (\mu \, {\rm amu})]$ 
is the one-dimensional isothermal sound speed in the gas, 
$\sigma_{\rm NT} = \Delta V_{\rm NT}/(8 \log{2})^{1/2}$ represents the non-thermal 
component of the velocity dispersion, and $\Delta V_{\rm NT} = 
[\Delta V_{\rm 13CO}^2 -8 \log{2} \, kT / m_{\rm 13CO}]^{1/2}$, where
$\Delta V_{\rm 13CO}$ is the observed linewidth.

%
%
\begin{figure}
\centering
\includegraphics[width=8cm]{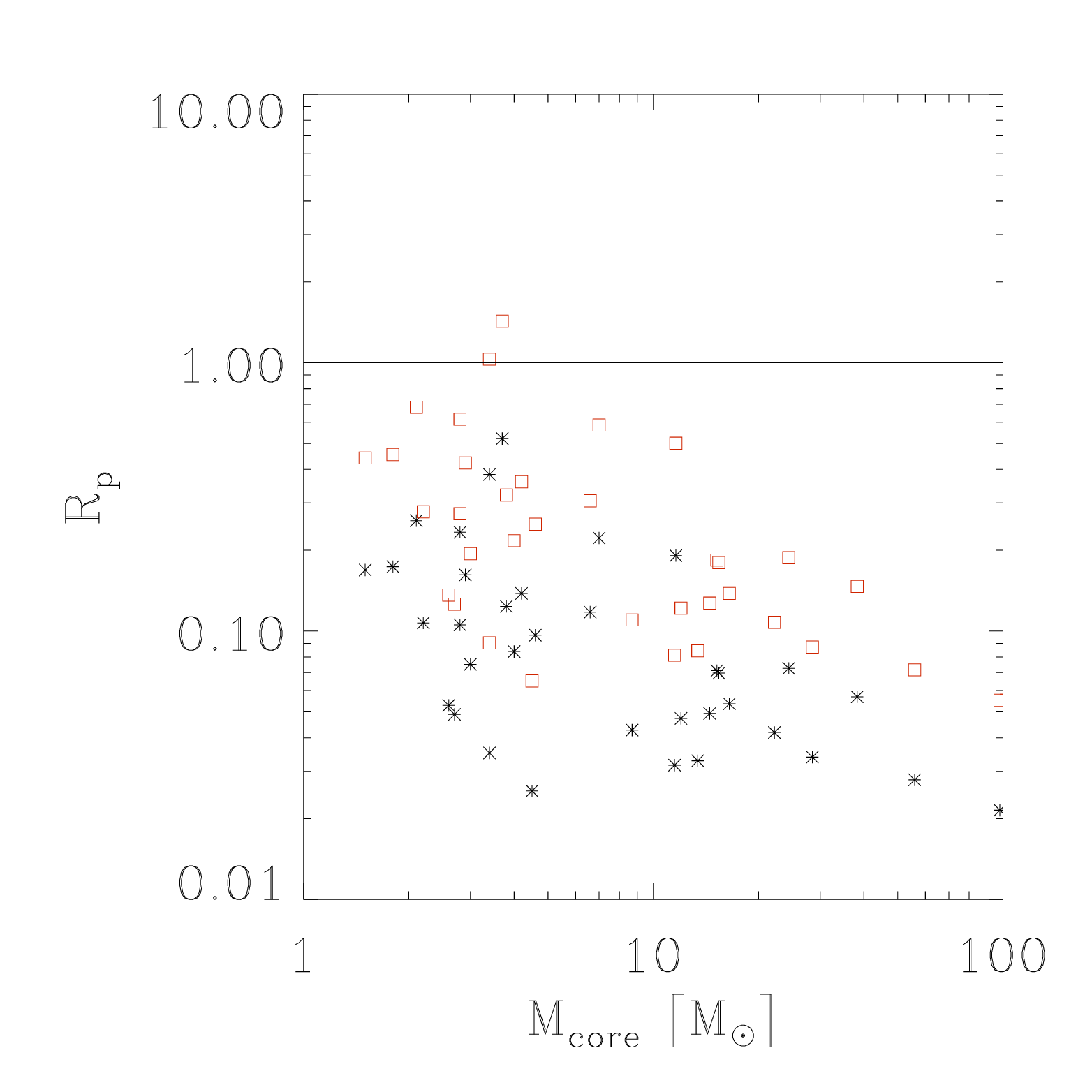}
\includegraphics[width=8cm]{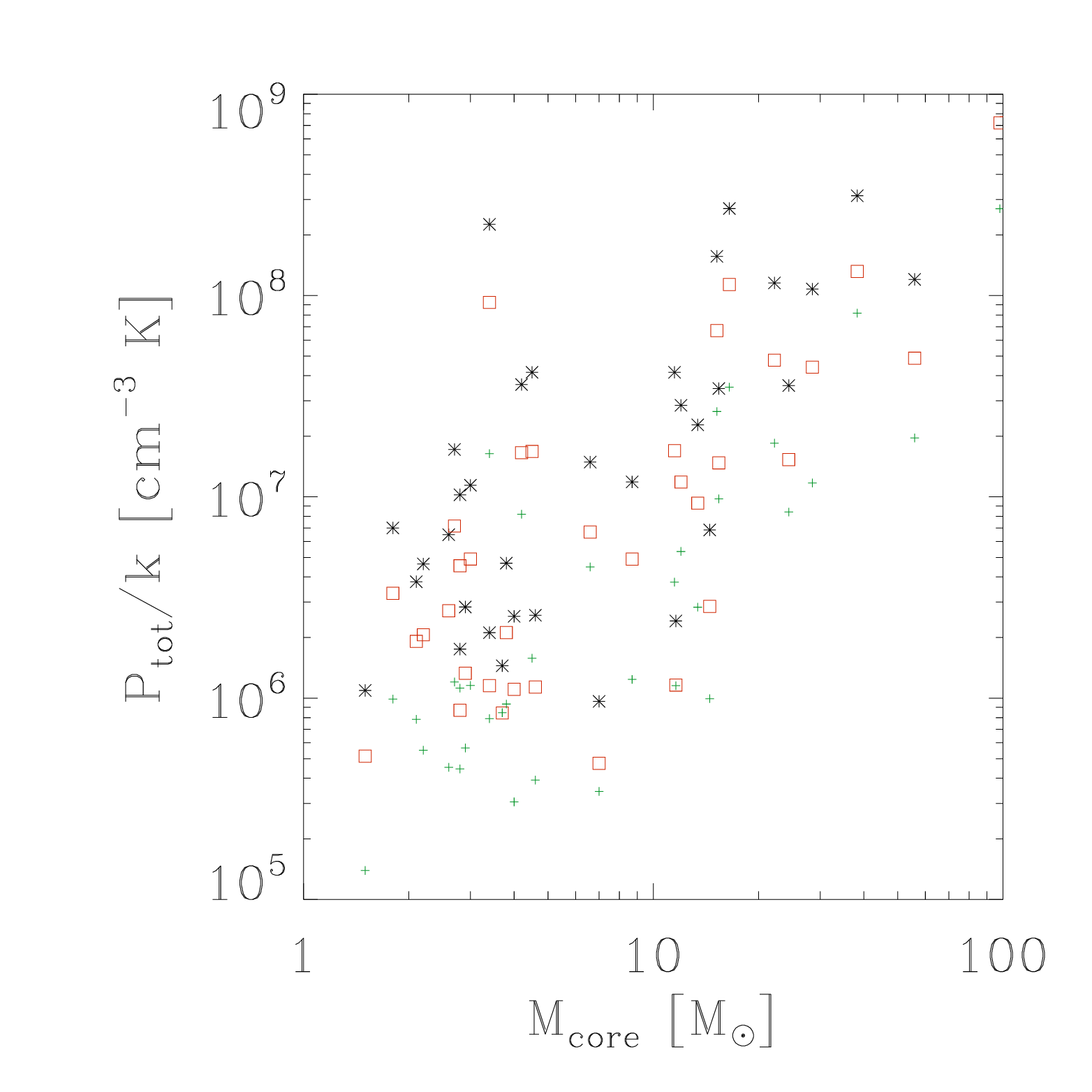}
\caption{
{\it Top.} Ratio of thermal to non-thermal pressure for the Vela-D cores vs. core mass.
{\it Bottom.} Total core internal gas pressure plotted as a function of core mass.
Symbols are as in Figure~\ref{fig:virial}. The small green ``+'' signs correspond to the 
pressure calculated using a Larson-type scaling law (see text).
  }
\label{fig:pressure}
\end{figure}

In the top panel of Figure~\ref{fig:pressure} we plot $R_p$ vs. core mass 
with and without the scale factor
for the linewidths (Section~\ref{sec:virial}). This figure shows that
all of the cores in Vela-D for which we have a measured $^{13}$CO$(2-1)$
linewidth are below the $R_p=1$ line, and thus the non-thermal pressure
would appear to exceed the thermal pressure. 
Figure~\ref{fig:pressure} also shows that the situation only marginally change 
if the linewidths are scaled with the 1.6 factor. Many more cores would 
have $R_p>1$ if the larger 2.8 scale factor from the Pipe cores were used.
In fact, \citet{lada2008} found in the Pipe cores that where both 
NH$_3$ and C$^{18}$O are observed, the non-thermal pressure derived from
the C$^{18}$O data is on average a factor of 4 larger than that measured
by the ammonia line. 

We then evaluate the {\it average} total internal pressure of each core in Vela-D
for which we have measured the $^{13}$CO$(2-1)$ linewidth. We note that this pressure will
overestimate the pressure at the edge of the cores. 
Including both thermal and non-thermal contributions, the pressure is given by:
\begin{equation}
P_{\rm tot} = \rho \, (a^2 + \sigma_{\rm NT}^2)
\label{eq:ptot}
\end{equation}
where $\rho$ is the mean density of a core.   
In the bottom panel of Figure~\ref{fig:pressure} we plot the 
quantity $P_{\rm tot}/k$ vs. core mass, 
where we note the high values of the pressure, due to the
relatively high density of the cores and to the large $^{13}$CO$(2-1)$ linewidths.
These values are consistent with the cores internal pressure estimated by \citet{massi07},
who also found that the external pressure to the cores is of order 
$P_{ext}/k \sim 5 \times 10^5$~cm$^{-3}$~K.
However, if the 1.6 scale factor for the linewidths is applied, the total internal pressure
would vary, for most cores (Figure~\ref{fig:pressure}), 
in the range $P_{\rm tot}/k \sim 5\times 10^5 - 4\times 10^7$~cm$^{-3}$~K, 
comparable to the ambient pressures characterizing other SF regions. 

In the lower panel of Figure~\ref{fig:pressure} it is also evident the trend of 
increasing $P_{\rm tot}/k$ for more massive cores. We find a trend of increasing density
with increasing core mass in Vela-D, but this trend is not enough to explain the
increase in $P_{\rm tot}/k$ with core mass over more than two orders of magnitude, as 
shown in Figure~\ref{fig:pressure}. Therefore, the higher value of $P_{\rm tot}/k$
in more massive cores must also be a consequence of the increased thermal and non-thermal
components of the internal pressure.  In particular, the top panel of Figure~\ref{fig:pressure} 
shows that the contribution of the non-thermal pressure becomes dominant for high-mass cores.

This is in fact consistent with the trend observed 
between pressure and mass when a simple scaling law \citep{larson1981} is used between the 
line width and the density, $\Delta V \propto \rho^{-1/2}$. According to \citet{adams1996} this implies a 
scaling relation between the linewidth and the mass of the core, which can be written as:
\begin{equation}
M_{\rm core} {\rm [M_\odot]} \simeq 466.5 \, (a_{\rm eff} {\rm [km \, s^{-1}]})^4
\end{equation}
this relation can also be written using Eq.~(\ref{eq:ptot}) for the ``effective'' sound speed,
$a_{\rm eff}^2 = P_{\rm tot} / \rho  $, as:
\begin{equation}
P_{\rm tot} {\rm [dyne \, cm^{-2}]} = 4.6\, 10^8 \, \rho {\rm [g \, cm^{-3}]} 
\, (M_{\rm core} {\rm [M_\odot]})^{1/2}
\label{eq:ptot_larson}
\end{equation}
which is plotted in the lower panel of Figure~\ref{fig:pressure}, in the usual 
units of $P_{\rm tot}/k$. We note that the model pressure reproduces the trend followed by
the data points, but for each specific point the pressure given by Eq.~(\ref{eq:ptot_larson})
generally underestimates the pressure given by Eq.~(\ref{eq:ptot}), though to a lesser extent
when the scale factor for the linewidths is applied.


\subsection{What Supports the Vela-D Cores?}
\label{sec:support}

We thus have two possible scenarios regarding the dynamical state of the Vela-D cores, 
if we consider the sub-sample
of cores within the area covered by  the $^{13}$CO$(2-1)$ map as a representative
sample of the whole Vela-D. If we do {\it not} apply the scale factor for the linewidths,
then almost all of the cores would appear to be gravitationally {\it unbound} 
(Figure~\ref{fig:virial}) and would have an internal pressure much higher
than the core external pressure (Figure~\ref{fig:pressure}). 
%
%
We note, however, that the average value of the core deconvolved radius to sound speed ratio is
$\langle R_{\rm dec}/a \rangle = (3.2\pm1.1)\times 10^5\,$yr, thus quite lower
compared to the estimated dynamical age, $\sim 1$ to $2\times 10^6\,$yr, 
of the filaments and stellar clusters in the Vela-D region (see \citealp{elia2007} 
and \citealp{massi10}). This would suggest that either the gravitationally unbound cores
are really transient structures 
(e.g., produced by the turbulence of the Vela-D region; see \citealp{elia2007} and \citealp{massi07}) 
and are thus constantly replenished, or they are externally supported.

As for the latter case, Figure~\ref{fig:pressure} shows that many of these cores
cannot be pressure-confined. Then the question arises of how the gravitationally unbound cores can be 
effectively confined against their own internal pressure to still form coherent structures. 
Because in this case self-gravity and surface pressure alone appear to be
insufficient to bind the cores,
a plausible source of external confinement could come from the presence of
static, helical magnetic fields within the filaments or clumps where the dense cores
have formed (see \citealp{fiege2000}). 

We can estimate the magnitude of the 
field strengths needed to produce
the external magnetic pressure which would add to the ambient gas pressure
to compensate the cores internal pressure, i.e. 
$P_B = B^2/(8\pi) = P_{\rm tot} - P_{\rm ext}$. In this way we find that the required 
$B$ would vary from a few tens to several hundreds of $\mu$G.
While these values are in agreement with measured magnitudes of $B$ in molecular clouds
(see, e.g., \citealp{basu2004} and references therein), we would have to assume that all
of this field is toroidal and constitutes the dominant component \citep{fiege2000}. 

On the other hand, we know that the $^{13}$CO$(2-1)$ linewidths are indeed
overestimating the core internal velocity dispersion, as discussed in the previous sections. 
Thus, if we do apply the approximate scale factor 
discussed in Section~\ref{sec:virial} then $\sim 30-50$\% (depending on the scale factor assumed) 
of the cores (mostly proto-stellar) 
would become gravitationally bound. The internal pressure of these cores 
still appears to be significantly higher than 
the pressure of the ambient medium, but because the cores would
be confined by self-gravity  the high surface pressures are of no
concern for the overall long-term stability of the cores.

The cores that are {\it not} gravitationally bound, and with internal pressures higher
than the surface pressure, would still need a source of external confinement, unless
they are transient structures, as discussed earlier. However,
we find that many ($\simeq 40$\%) of these unbound cores (both starless and proto-stellar) 
would have an internal pressure $P_{\rm tot}/k \la 2\times 10^6$, which is a factor of 4 
higher than the estimated average external pressure in Vela-D, 
$P_{\rm ext}/k \sim 5 \times 10^5$~cm$^{-3}$~K \citep{massi07},
thus partially reducing the imbalance between internal and external pressures.

\section{SUMMARY AND CONCLUSIONS}
\label{sec:concl}

In this paper we have analyzed the dynamical state of the population of the dense 
and cold cores detected by BLAST in the Vela-D molecular cloud, utilizing the 
sensitive maps at 250, 350 and 500\,\micron\ 
obtained by BLAST during its 2006 LDB flight from Antarctica.
We can summarize the results of this paper as follows, although
some of our conclusions are still dependent on an exact measurement of the internal
velocity dispersion of the cores: 

1. The median $M_{\rm core}/M_{\rm BE} = 0.9$ for all sources in Vela-D,
i.e. very near the critical value, assuming thermal support only.
If we restrict ourselves to
mass ranges for which we are progressively more complete,
our sources exceed their Bonnor-Ebert mass by a median factor of $M_{\rm core}/M_{\rm BE} = 1.5 - 3.3$.

2. If the BLAST cores were purely isothermal structures, then almost all would
be either confined by the external gas pressure, or by their gravity. 
According to the virial analysis of the cores with an unambiguous 
line-of-sight $^{13}$CO(2-1) spectrum, almost none of them would appear to be
gravitationally bound and would also be dominated by non-thermal pressure.
However, we think this is likely a consequence of the $^{13}$CO(2-1) 
linewidths overestimating the core internal velocity dispersion. 
Despite the uncertainties involved in this procedure,
we have attempted to correct for this effect and found that $\sim 30-50$\%
of the cores would then turn out to be gravitationally bound.

3.  The average total internal pressure of the cores 
appear to be higher than the ambient external pressure.  
If most cores were indeed gravitationally {\it unbound}
then self-gravity and surface pressure would be insufficient to bind the
cores and form coherent structures. After application of the
scale factor to the $^{13}$CO(2-1) linewidths, a significant fraction of the cores 
(mostly proto-stellar) would actually become gravitationally bound, and thus they would 
not need to be pressure-confined.  

4. The cores that result gravitationally unbound even after application of 
the scale factor to the $^{13}$CO(2-1) linewidths,
would still need an additional source of external confinement. If the
confinement mechanism were an external magnetic pressure, the required
$B$ would vary from a few tens to several hundreds of $\mu$G.
Measurements of the BLAST cores velocity dispersion is necessary to allow a 
more accurate dynamical analysis.

5. Alternatively, the gravitationally unbound cores could be transient structures,
continuously produced in the region by turbulence.

\acknowledgments
We acknowledge the support of NASA through grant numbers NAG5-12785,
NAG5-13301, and NNGO-6GI11G, the NSF Office of Polar Programs, the Canadian
Space Agency, the Natural Sciences and Engineering Research Council (NSERC) of
Canada, and the UK Science and Technology Facilities Council (STFC).
Support for this work was provided by NASA through an award issued by JPL/Caltech

\bibliographystyle{apj}
\bibliography{apj-jour,refs}

\begin{thebibliography}{27}
\expandafter\ifx\csname natexlab\endcsname\relax\def\natexlab#1{#1}\fi

\bibitem[{{Adams} \& {Fatuzzo}(1996)}]{adams1996}
{Adams}, F.~C. \& {Fatuzzo}, M. 1996, \apj, 464, 256

\bibitem[{{Andre} {et~al.}(2000){Andre}, {Ward-Thompson}, \&
  {Barsony}}]{andre2000}
{Andre}, P., {Ward-Thompson}, D., \& {Barsony}, M. 2000, Protostars and Planets
  IV, 59

\bibitem[{{Basu}(2004)}]{basu2004}
{Basu}, S. 2004, in Young Local Universe, Proceedings of XXXIXth Rencontres de
  Moriond, ed. A.~{Chalabaev}, T.~{Fukui}, T.~{Montmerle}, \&
  J.~{Tran-Thanh-Van}

\bibitem[{{Elia}(2007)}]{elia2007}
{Elia}, D.~{\it et al.}. 2007, \apj, 655, 316

\bibitem[{{Enoch} {et~al.}(2008){Enoch}, {Evans}, {Sargent}, {Glenn},
  Rosolowsky, \& Myers}]{enoch2008}
{Enoch}, M.~L., {Evans}, II, N.~J., {Sargent}, A.~I., {Glenn}, J., Rosolowsky,
  E., \& Myers, P. 2008, \apj, 684, 1240

\bibitem[{{Enoch} {et~al.}(2007){Enoch}, {Glenn}, {Evans}, {Sargent}, {Young},
  \& {Huard}}]{enoch2007}
{Enoch}, M.~L., {Glenn}, J., {Evans}, II, N.~J., {Sargent}, A.~I., {Young},
  K.~E., \& {Huard}, T.~L. 2007, \apj, 666, 982

\bibitem[{{Fiege} \& {Pudritz}(2000)}]{fiege2000}
{Fiege}, J.~D. \& {Pudritz}, R.~E. 2000, \mnras, 311, 85

\bibitem[{{Giannini}(2007)}]{gianni07}
{Giannini}, T.~{\it et al.}. 2007, \apj, 671, 470

\bibitem[{{Jijina} {et~al.}(1999){Jijina}, {Myers}, \& {Adams}}]{jijina1999}
{Jijina}, J., {Myers}, P.~C., \& {Adams}, F.~C. 1999, \apjs, 125, 161

\bibitem[{{Lada} {et~al.}(2008){Lada}, {Muench}, {Rathborne}, {Alves}, \&
  {Lombardi}}]{lada2008}
{Lada}, C.~J., {Muench}, A.~A., {Rathborne}, J., {Alves}, J.~F., \& {Lombardi},
  M. 2008, \apj, 672, 410

\bibitem[{{Larson}(1981)}]{larson1981}
{Larson}, R.~B. 1981, \mnras, 194, 809

\bibitem[{{Liseau} {et~al.}(1992){Liseau}, {Lorenzetti}, {Nisini}, {Spinoglio},
  \& {Moneti}}]{Lis92}
{Liseau}, R., {Lorenzetti}, D., {Nisini}, B., {Spinoglio}, L., \& {Moneti}, A.
  1992, \aap, 265, 577

\bibitem[{MacLaren {et~al.}(1988)MacLaren, Richardson, \&
  Wolfendale}]{maclaren1988}
MacLaren, I., Richardson, K.~M., \& Wolfendale, A.~W. 1988, \apj, 333, 821

\bibitem[{{Massi} {et~al.}(2007){Massi}, {De Luca}, {Elia}, {Giannini},
  {Lorenzetti}, \& {Nisini}}]{massi07}
{Massi}, F., {De Luca}, M., {Elia}, D., {Giannini}, T., {Lorenzetti}, D., \&
  {Nisini}, B. 2007, \aap, 466, 1013

\bibitem[{{Massi} {et~al.}(2010){Massi}, {di Carlo}, {Codella}, {Testi},
  {Vanzi}, \& {Gomes}}]{massi10}
{Massi}, F., {di Carlo}, E., {Codella}, C., {Testi}, L., {Vanzi}, L., \&
  {Gomes}, J.~I. 2010, \aap, 516

\bibitem[{{McKee} \& {Ostriker}(2007)}]{mckee2007}
{McKee}, C.~F. \& {Ostriker}, E.~C. 2007, \araa, 45, 565

\bibitem[{{Muench} {et~al.}(2007){Muench}, {Lada}, {Rathborne}, {Alves}, \&
  {Lombardi}}]{muench2007}
{Muench}, A.~A., {Lada}, C.~J., {Rathborne}, J.~M., {Alves}, J.~F., \&
  {Lombardi}, M. 2007, \apj, 671, 1820

\bibitem[{{Murphy} \& {May}(1991)}]{mur:may}
{Murphy}, D.~C. \& {May}, J. 1991, \aap, 247, 202

\bibitem[{{Netterfield}(2009)}]{netterfield2009}
{Netterfield}, C.~B.~{\it et al.}. 2009, \apj, 707, 1824

\bibitem[{{Olmi}(2009)}]{olmi2009}
{Olmi}, L.~{\it et al.}. 2009, \apj, 707, 1836

\bibitem[{{Onishi}(1999)}]{onishi1999}
{Onishi}, T.~{\it et al.}. 1999, \pasj, 51, 871

\bibitem[{{Pascale}(2008)}]{pascale2008}
{Pascale}, E.~{\it et al.}. 2008, \apj, 681, 400

\bibitem[{{Stahler} \& {Palla}(2004)}]{sp2004}
{Stahler}, S.~W. \& {Palla}, F. 2004, in The Formation of Stars, ed. WILEY-VCH

\bibitem[{{Vilas-Boas} {et~al.}(2000){Vilas-Boas}, {Myers}, \&
  {Fuller}}]{vilasboas2000}
{Vilas-Boas}, J.~W.~S., {Myers}, P.~C., \& {Fuller}, G.~A. 2000, \apj, 532,
  1038

\bibitem[{{Ward-Thompson} {et~al.}(1999){Ward-Thompson}, {Motte}, \&
  {Andr\'e}}]{ward1999}
{Ward-Thompson}, D., {Motte}, F., \& {Andr\'e}, P. 1999, \mnras, 305, 143

\bibitem[{{Williams} {et~al.}(1994){Williams}, {de Geus}, \&
  {Blitz}}]{williams1994}
{Williams}, J.~P., {de Geus}, E.~J., \& {Blitz}, L. 1994, \apj, 428, 693

\bibitem[{{Young} {et~al.}(2003){Young}, {Shirley}, {Evans}, \&
  {Rawlings}}]{young2003}
{Young}, C., {Shirley}, Y., {Evans}, N.~I., \& {Rawlings}, J. 2003, \apjl, 145,
  111

\end{thebibliography}

\end{document}